\documentclass[aps,prb,twocolumn,superscriptaddress]{revtex4-1}
\usepackage{amsmath}
\usepackage{bm}
\usepackage{graphicx}
\usepackage{amsfonts}
\usepackage{amssymb}

\newcommand{\be}{\begin{equation}}
\newcommand{\ee}{  \end{equation}}
\newcommand{\ba}{\begin{eqnarray}}
\newcommand{\ea}{  \end{eqnarray}}
\newcommand{\bi}{\begin{itemize}}
\newcommand{\ei}{  \end{itemize}}

\begin{document}
\title{Fano-Kondo and the Kondo-box regimes crossover  in a quantum dot coupled to a quantum box}
\author{Victor .M. Apel}
\affiliation{Departamento de F\'{\i}sica, Universidad Cat\'{o}lica
del Norte, Casilla 1280, Antofagasta, Chile}
\author{Pedro A. Orellana}
\author{Monica Pacheco}
\affiliation{Departamento de F\'{\i }sica, Universidad T\'{e}cnica
F. Santa Maria, Casilla Postal 110 V, Valparaiso, Chile}
\author{Enrique V. Anda}
\affiliation{Departamento de F\'{\i }sica, P. U. Cat\'{o}lica do
Rio de Janeiro, C.P. 38071-970, Rio de Janeiro, RJ, Brazil}

\begin{abstract}

In this work,we study the Kondo effect of a quantum dot (QD) connected to leads and to a discrete set of one particle states provided by a quantum box represented by a quantum ring (QR) pierced by a magnetic flux side attached to the QD.  The interplay between the bulk Kondo effect and the so called Kondo box regime is studied.  In this system the QR  energies can be continuously modified by the application of the magnetic field. The crossover between these two regimes is analyzed by changing the connection of the QD to the QR from the weak to the strong coupling regime.
In the weak coupling regime, the differential conductance develops a sequence of
Fano-Kondo antiresonances due to destructive interferences between the discrete quantum ring levels and the conducting Kondo channel provided by the leads. In the strong coupling regime the differential conductance has very sharp resonances when one of the Kondo discrete sub-level characterizing the Kondo box is tunned by the applied potential.  The conductance, the current fluctuations and the Fano coefficient result to be the relevant physical magnitudes to be analyzed to reveal the physical properties of these two Kondo regimes and the crossover region between them. The results were obtained by using the Slave Boson Mean Field Theory (SBMFT).
\end{abstract}
\maketitle

\section{Introduction}
Electronic transport through nanoscopic systems, as quantum dots (QDs) has been
extensively studied in the last decade
\cite{GR88,GSMAMK98,Kouwenhoven,GGHK00}.  The QDs allow the systematic study of
quantum-coherent effects as Kondo, Fano and
Aharonov-Bohm effects due to the possibility of continuous tuning the
relevant parameters governing the properties of these regimes,
in equilibrium and nonequilibrium
situations\cite{GR88,GGHK00,fano,Kobayashi1,Kobayashi2,Wu}.

Kondo effect in QDs dominates the electronic transport properties due to the strong many-body
correlations between the conduction band electron spins in the leads
and the localized spin in the QD, when the temperature $T$ is below the Kondo temperature $T_{k}$ \cite{GSMAMK98,K64}.
The main
signature of the Kondo state in nanosystems as a QD connected to two leads is the
enhancement of the
conductance below $T_{k}$ up to the unitary limit ($2e^{2}/h$)
\cite{GSMAMK98}. In
this configuration electrons transmitted from one electrode to
the other necessarily pass through the QD. It is presently known
that coupled quantum dots exhibit the electronic counterpart of
the Fano and Dicke effects and that they can be controlled by the
magnetic flux and quantum-dot asymmetry
\cite{Lara,Orellana,Davi,Trocha}. The emergence of Fano resonances requires
 two scattering channels: a discrete level and
a broad continuum band \cite{fano}. Present understanding of
electron transport properties of quantum dots is based mainly on
direct current measurements. However, additional information can
be obtained from fluctuations of the current \cite{blanter,
bing1,Reznikov,Nakamura,Nakamura1,Hashisaka,Egger,Delattre,yeyati}.
Electronic current through any conductor fluctuates with time reflecting
charge granularity. This phenomenon is referred in the literature as shot noise.
It has been demonstrated that
electron shot noise provides a useful tool to detect the role played by
electron coherence and Coulomb interactions and correlation in electronic
transport through quantum dots\cite{bing1,yeyati}. Besides it provides information about current
fluctuations that cannot be extracted from the average current
alone \cite{Egger,Delattre}.

There have been theoretical\cite{Kbox,Kbox2,Kbox3} and experimental studies\cite{KboxExp1,KboxExp2} of the Kondo phenomena when the QD is connected to a quantum box, where the uncorrelated conducting electrons are described by a series of discrete energy levels. The finite system density of states consists of a series of peaks separated by an energy $\Delta$ inversely proportional to the number of sites $N$ and, in a tight binding representation of the discrete system, proportionally to the hopping matrix elements $v$ that connects the sites. The universal exponential dependence with the system parameters of the Kondo temperature is established in this case in the limit when the energy separation of the discrete levels of the quantum box is much less than the characteristic Kondo temperature that this system would have if the impurity were connected to a continuum conduction band. The behavior of a Kondo box, from the discrete to the continuum limit when the size of the box tends to infinite, has been studied\cite{Kbox,Simon02_03}. For the case of a QD connected to a quantum box able to be represented in the Fermi region by only one semi-occupied level, the energy of the singlet ground state is less than the energy of the excited triplet state by an amount that can be associated to  a $T_k$ of this discrete system. This $T_k$ can be expanded in powers of the parameter $\gamma=t^2/\varepsilon_{0}$
where $t$ is the non diagonal hopping matrix elements connecting the QD to the quantum box and $\varepsilon_0$ is the energy of the local QD level\cite{Fulde}.

In the present work, we investigate the transport properties of a
strongly correlated QD attached to two leads and to a quantum ring (QR) pierced by a magnetic flux, that in fact acts as a quantum box, which energies can be continuously modified by the application of the magnetic field. This system is  conditioned by the interplay between two different Kondo effects, the bulk Kondo regime that results from the connection of the QD to the leads and the Kondo box regime due to the interaction between the QD and the ring. Manipulating the parameters, the system presents unexplored very interesting crossover behavior. A schematic model of the structure proposed is shown in Fig. \ref{scheme}. We call the attention to the fact that this system possesses similarities to the one proposed to study the properties of a two channel Kondo system\cite{KboxExp1}. However, the physics we are analyzing is completely different because our system does not have two independent channels as in this case  the electrons can hop from the continuous to the discrete reservoir without any restrictions.There is no Coulomb correlation among the electrons within the QR to impede the free entrance of an electron when one of the discrete levels is in the nearby of the Fermi energy\cite{KboxExp1}.

The crossover between the continuous Kondo and the Kondo box regimes can be studied by manipulating the connection of the QD to the leads and to the QR. The conductance, the current fluctuations and the Fano coefficient result to be the relevant physical magnitudes to be analyzed in the parameter space to reveal the physical properties of these two Kondo regimes and the crossover region between them.
The physics of the QR acting as a quantum box and the crossover between the two regimes depend upon the spacing of the QR states in comparison with the Kondo temperature and on the quantum box being at resonance (off-resonance) when one of its states is (is not) at the neighborhood of the Fermi level. This last condition can be changed continuously by simply modifying the magnetic flux threading the ring.

The results are obtained using the SBMFT \cite{Wu,SB,Newns} approach that is able to adequately describe the properties of the Kondo regime at very low energies, far apart from the mix-valence regime.

The paper is organized as follows. In Section \ref{Model}, we present the model, the Hamiltonian we used to study it, the central concepts regarding the SBMFA and the derivation of the analytical expression for the current $I$ and  shot noise $S$. In Section \ref{Numeral Result}, we briefly discuss the results for the LDOS and the transport properties corresponding to the different regimes of the system. Finally, we elaborate a summary of the paper in Section \ref{Summary}.

\section{Model}\label{Model}

\begin{figure}
\vspace*{1cm} \centering
\rotatebox{90}{\scalebox{0.5}{\includegraphics{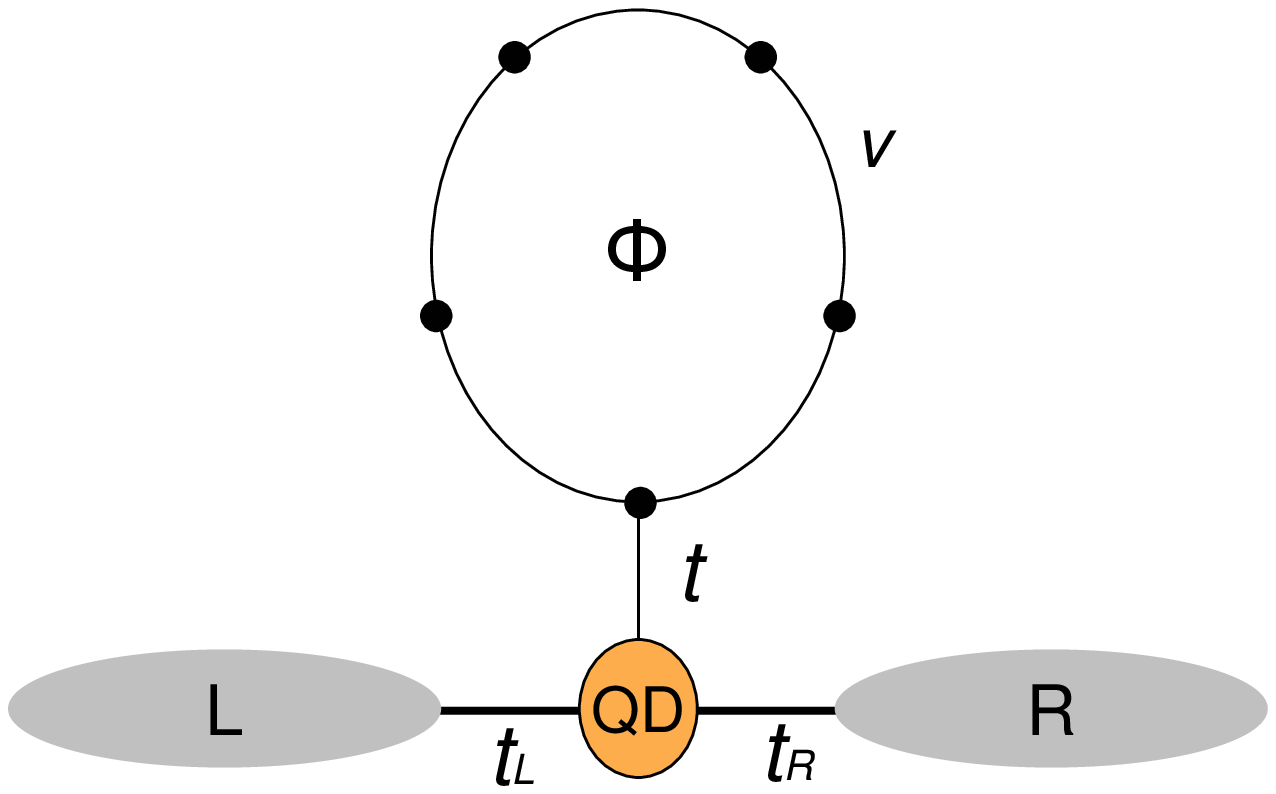}}}
\caption{ Schematic diagram of quantum-dot  attached to two leads and coupled to a quantum ring (quantum box)  pierced by a magnetic flux.} \label{scheme}
\end{figure}

We consider QD with e-e interaction, connected with two leads and
with $N$ sites QR. The Hamiltonian of the system outlined in Fig.
\ref{scheme} can be written as,
\begin{equation*}
H=H_{L}+H_{QR}+H_{QD}+H_{L-QD}+H_{QD-QR}
\end{equation*}
where the different sub-Hamiltonian $H_\beta$ are given by,
\begin{eqnarray}
H_{L} &=&\sum_{k_{\alpha}\sigma,\alpha}\varepsilon_{k_{\alpha}}c^{\dag}_{k_{\alpha}\sigma}c_{k_{\alpha}\sigma}+hc\nonumber\\
H_{QR}&=&\sum_{m\sigma}\varepsilon_{m}(\Phi)a^{\dag}_{m\sigma} a_{m\sigma} \nonumber\\
H_{QD}&=&\varepsilon_{0}(n_{\uparrow}+n_{\downarrow})+Un_{\uparrow}n_{\downarrow}\nonumber\\
H_{L-QD}&=&\sum_{k_\alpha, \alpha}t_{\alpha}c^{\dag}_{k_{\alpha}\sigma}d_{\sigma}+hc\nonumber\\
H_{QD-QR}&=&\frac{t}{\sqrt{N}}\sum_{m,\sigma}a^{\dag}_{m\sigma}d_{\sigma}+hc
\end{eqnarray}
where $\alpha=L,R$,$\varepsilon_{m}(\Phi)=-2v\cos(\frac{2\pi}{N}(\Phi/\Phi_{0}+m))$ ($m=1,..,m$) and $\Phi_{0}=h/e$ is the quantum of flux. We consider the density of states describing the left (right) lead $\rho_{L(R)}$ as being constant and equal to $1/D$, where D is the
lead bandwidth. The coupling strength between the QD and the leads is given by
 $\Gamma_{L(R)}=\pi\rho_{L(R)}t_{L(R)}^{2}$.

In the wide band limit ($D>>1$) and for infinite Coulomb repulsion
($U\rightarrow\infty$), $T_{k}=De^{-\pi|\varepsilon_{0}|/\Gamma}$. It is necessary that $T<T_{k}$ for the system to have a Kondo
behavior.  When the temperature $T>T_{k}$,
the spin correlations
between the localized electron and the conduction electrons are eliminated by thermal fluctuation,
driving the system out of Kondo regime. The Kondo ground state corresponds to a situation in which the QD is occupied by a one electron, which requires $\varepsilon_{0}<-\Gamma
$ and $(\varepsilon_{0}+U)>\Gamma$. Otherwise, charge fluctuations are significant and the system is in a mixed valence regime.

The intra-dot Coulomb interaction U is not a relevant parameter for the Kondo effect if the QD is in the one electron Coulomb blockade regime. As a consequence, it is interesting to assume the $U\rightarrow\infty$ limit as it simplifies the self-consistent calculations because the QD in this case has only three available states: the empty state or the spin up or down electron occupied states\cite{Newns}. It is convenient to rewrite the Hamiltonian $H$,
projecting out the double QD occupancies. In order to
do this, we use the slave boson approach \cite{SB}. We define the
operators $f_{\sigma}^{\dag}$ ($f_{\sigma}$) that creates
(destroys) the pseudo fermion of $\sigma$ spin in  the QD and
$b^{\dag}$ ($b$) that creates (destroys) an empty pseudo bose state
and impose a constraint that explicitly eliminates the double occupied state,
\begin{eqnarray}
b^{\dag}b+\sum_{\sigma}f^{\dag}_{\sigma}f_{\sigma}=\textbf{1}
.\label{constraint}
\end{eqnarray}

The use of an auxiliary field $\lambda$ which acts as a Lagrange
multiplier permits to impose the condition Eq.\ref{constraint}. The Hamiltonian takes the form
\begin{eqnarray}
H&&=\sum_{k_{\alpha}\sigma, \alpha}\varepsilon_{k_{\alpha}}c^{\dag}_{k_{\alpha}\sigma}c_{k_{\alpha}\sigma}+\sum_{\sigma}
\varepsilon_{0}f_{\sigma}^{\dag}f_{\sigma}\nonumber\\
&&+\sum_{m}\varepsilon_{m}(\Phi)a^{\dag}_{m\sigma}a_{m\sigma}+\frac{1}{\sqrt{2}}\sum_{k_{\alpha}\sigma,
\alpha=L,R}t_{\alpha}c^{\dag}_{k_{\alpha}\sigma}b^{\dag}f_{\sigma}\nonumber\\
&&+
\frac{t}{\sqrt{2N}}\sum_{m}a^{\dag}_{m\sigma}b^{\dag}_{\sigma}f_{\sigma}+\lambda(b^{\dag}b+\sum_{\sigma}f^{\dag}_{\sigma}f_{\sigma}-1)
\label{bo}
\end{eqnarray}

Within the SBMFT approach
\cite{Newns},  we replace the
bosonic operator by its expectation
value $\langle b\rangle=\sqrt{2} \tilde{b}$ and assume that
$\langle b^{\dag}b\rangle\simeq\langle b\rangle^{2}=2 \tilde{b}
$. At temperature $T=0$, SBMFT describes quite well the Kondo
regime characterized by strong spin fluctuation, but it is not
reliable in the charged fluctuating mixed valence region. This confines our analysis to
$\varepsilon_{0}/(\Gamma_{L}+\Gamma_{R})<-0.5$.  Within this formalism, the Hamiltonian becomes a one-body effective one, $\tilde{H}$.
\begin{eqnarray}
\tilde{H}&&=\sum_{k\sigma\alpha}\varepsilon_{k_\alpha}c^{\dag}_{k_{\alpha}\sigma}c_{k_{\alpha}\sigma}+\sum_{\sigma}
\tilde{\varepsilon_{0}}f_{\sigma}^{\dag}f_{\sigma}+\sum_{m}\varepsilon_{m}(\Phi)a^{\dag}_{m\sigma}a_{m\sigma}\nonumber\\
&+&\sum_{k_{\alpha}\sigma,
\alpha}(\tilde{t_{\alpha}}c^{\dag}_{k_{\alpha}\sigma}f_{\sigma}+h.c.)\nonumber+\frac{\tilde{t}}{\sqrt{N}}\sum_{m,\sigma}(a^{\dag}_{m\sigma}f_{\sigma}+h.c.)\nonumber\\
&+&\lambda(2\tilde{b}^{2}-1)
\label{effective}
\end{eqnarray}
where $\tilde{t}_{L(R)}=t_{L(R)}\tilde{b}$, $\tilde{t}=t\tilde{b}$
and $\tilde{\varepsilon_{0}}=\varepsilon_{0}+\lambda$ are
renormalized parameters. By minimizing the expectation value of $\tilde{H}$, $\langle \tilde{H} \rangle$ with respect to $\tilde{b}$ and $\lambda$, incorporating the constraint (Eq.\ref{constraint}), we obtain,
\begin{eqnarray}
\tilde{b}^{2}=\frac{1}{2}-\frac{1}{2}\sum_{\sigma} \langle
f_{\sigma}^{\dag}f_{\sigma}\rangle.
\label{expectation}
\end{eqnarray}

We  determine self-consistently the parameters $\tilde{b}$ and $\lambda$ in terms of the Fourier component of QDs lesser Green's function\cite{Wu,Langreth}
$G_{d,\sigma}^{<}(\omega)=i\langle
f_{\sigma}^{\dag}f_{\sigma}\rangle_\omega$,
\begin{eqnarray}
\tilde{b}^{2}&=&\frac{1}{2}-\frac{i}{4\pi}\sum_{\sigma}\int_{-D}^D
G_{d,\sigma}^{<}(\omega)d\omega\nonumber \\
\lambda\tilde{b}^{2}&=&\frac{i}{4\pi}\sum_{\sigma}\int_{-D}^D
(\omega-\tilde{\varepsilon}_{0\sigma})G_{d,\sigma}^{<}(\omega)d\omega
\label{selfconsistent}.
\end{eqnarray}

The function $G_{d,\sigma}^{<}$ and the retarded Green's function
$G_{d,\sigma}^{r}$
 are obtained using Langreth analytic continuation and Dyson equation respectively
 \cite{Jauho,Langreth}
\begin{eqnarray}
G_{d,\sigma}^{<}(\omega)&=&2\,i\frac{f_{L}(\omega)\tilde{\Gamma}_{L}+f_{R}(\omega)\tilde{\Gamma}_{R}}{(\omega-\tilde{\varepsilon}_{0}-\frac{Q(\omega)}{\sqrt{N}})^{2}+(\tilde{\Gamma}_{L}+\tilde{\Gamma}_{R})^{2}}
\nonumber
\\G_{d,\sigma}^{r}(\omega)&=&\frac{1}{\omega-\tilde{\varepsilon}_{0}-\frac{Q(\omega)}{\sqrt{N}}+i(\tilde{\Gamma}_{L}+\tilde{\Gamma}_{R})}
\label{G0ret}
\end{eqnarray}
where $Q(\omega)=\sum_{m}\tilde{t}^{2}/(\omega-\varepsilon_{m}(\Phi))$ and
 $\tilde{\Gamma}_{L(R)}=\tilde{b}^{2}\Gamma_{L(R)}$.

 In order to characterize the different regimes of the system and its transport properties, it is important to calculate the DOS at the QD. It is given by the expression,

\begin{equation}
\rho_{QD}(\omega)=-\frac{1}{\pi}ImG^{r}_{d,\sigma}(\omega).
\label{DOS}
\end{equation}

On the other hand, the transmission probability $T(\omega,V)$ can be written,

\begin{eqnarray}
T(\omega,V)=4\frac{\tilde{\Gamma}_{L}\tilde{\Gamma}_{R}}{\tilde{\Gamma}_{L}+\tilde{\Gamma}_{R}}Im\big[G^{r}_{d,\sigma}(\omega)\big].
\end{eqnarray}

Note that the transmission depends on the applied bias $V$ as it appears in the self-consistent equations
Eq.\ref{selfconsistent}. The
transport properties are studied at temperature $T=0$. We calculate the current $I$ transmitted through the
QD, the shot noise $S$ and the Fano factor $FF=S/2eI$.  The current $I$ and the shot noise $S$ are given by, \cite{Meir,Lopez},

\begin{eqnarray}
I&=&\frac{2e}{h}\int_{-eV/2}^{eV/2} T(\omega,V)d\omega,\nonumber\\
S&=&\frac{4e^{2}}{h}\int_{-eV/2}^{eV/2}
T(\omega,V)[1-T(\omega,V)]d\omega.
\label{current-shotnoise}
\end{eqnarray}.

\section{Numerical Result}\label{Numeral Result}

In this section we discuss the transport properties at zero
temperature $T=0$. In what follows, we will consider $\Gamma_L=\Gamma_R=\Gamma$ as the energy unit and $E_{F}=0$. We have taken the band-with to be $D=35$ and set the QD
energy level at $\varepsilon_{0}=-3$. For these values
$T_{K}=1.4\times 10^{-3}$ and $\sum_{\sigma}\langle
f^{\dag}_{\sigma}f_{\sigma}\rangle=1-T_{K}\approx1$.

As already discussed in the introduction, in the Kondo box regime, the Kondo temperature of the equivalent continuous system is less than, or of the order of the energy separation of the quantum box states. In this regime, supposing an even number of electrons in the system, the energy difference between the ground singlet Kondo like state and the first excited triplet state is given by $4t^2/\varepsilon_{0}$\cite{Fulde}, which permits this value to be associated to the Kondo temperature of the quantum box. As a consequence, to characterize the regime of the system it is convenient to compare this typical energy with the Kondo temperature that results from the connection of the QD with the leads. This permits us to define a weak (strong) coupling regime when
$t<\xi$
($t>\xi)$, with $\xi\equiv\sqrt{\varepsilon_{0}T_{K}/2}$.

\begin{figure}[tbp] \centering
\rotatebox{0}{\scalebox{0.45}{\includegraphics{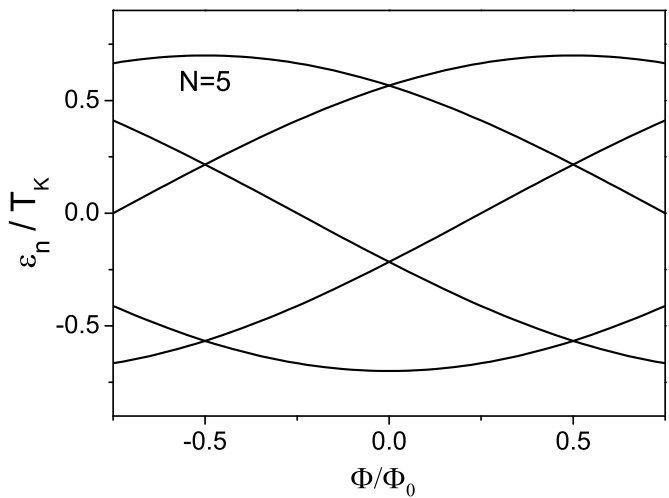}}}
\caption{Level spectrum of disconnected QR ($t=0$), as a function
of magnetic flux $\Phi$ for QR with $N=5$ sites.}
\label{spectrum}
\end{figure}

As far as the quantum box is concerned the transport properties are affected by the QR energies $\varepsilon_{n}(\Phi)$ and its proximity to the Fermi level. The values of $\varepsilon_{n}(\Phi)$ are  controlled  by $\Phi$, as shown in  Fig. \ref{spectrum},  while the spacing of the levels is dependent upon the magnitude $v/N$, being $N$ the number of sites of the ring and $v$ the hopping matrix element connecting them. As for numerical reasons it is better to take a small value of $N$, we  arbitrary assume $N=5$ and a value of $v$, $v=0.35T_k$, such that the few QR state energies are within the width of the Kondo peak of the QD attached to the leads and non connected to the ring. This is the condition for the properties of the Kondo regime, derived from the coexistence of a continuous and a discrete bath, to appear. Other possible states of a longer ring, far apart from the Fermi level, are not relevant for the physics we analyzed. We assume various magnetic fluxes $\Phi$ so as to manipulate the ring states relative to the Fermi level, studying in particular the situations when one state of the quantum ring is half occupied or the nearest states to the Fermi level are double or non-occupied.

\begin{figure}[tbp]
\centering
\rotatebox{0}{\scalebox{0.7}{\includegraphics{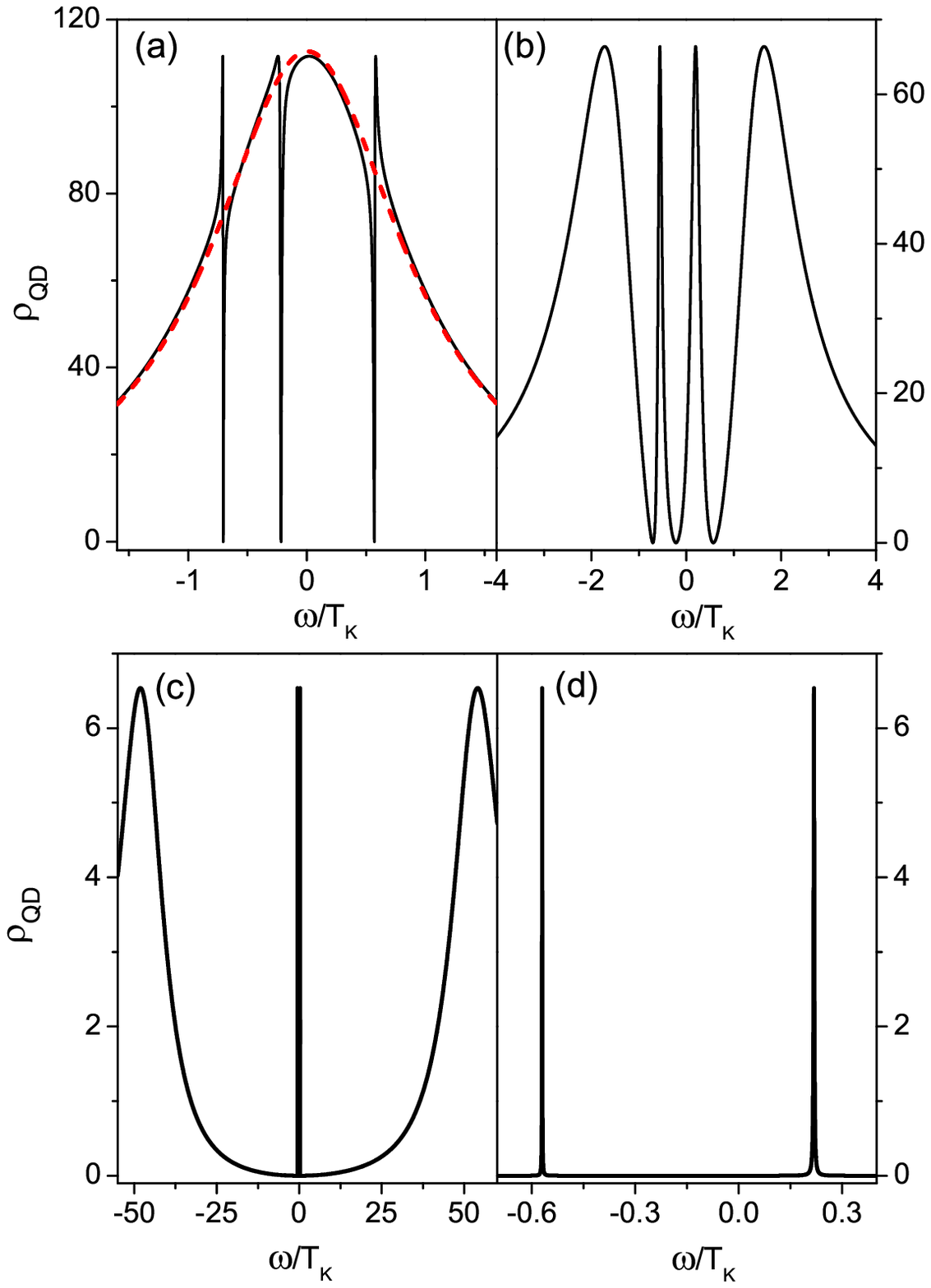}}}
\caption{(Color online) Density of states as a function of energy with magnetic flux $\Phi=0$. In the panel (a)   $t=0.1\, \xi$ (solid black line) and  for disconnected QR, $t=0$ (red dashed line). In the panel (b)
$t=\xi$ and in the panel (c) $t=10\,\xi$. In the panel (d) it is shown in detail, the two central Lorentzian peaks for  $t=10\,\xi$.}
\label{figDOS}
\end{figure}

\begin{figure}[tbp]
\centering
\rotatebox{0}{\scalebox{0.55}{\includegraphics{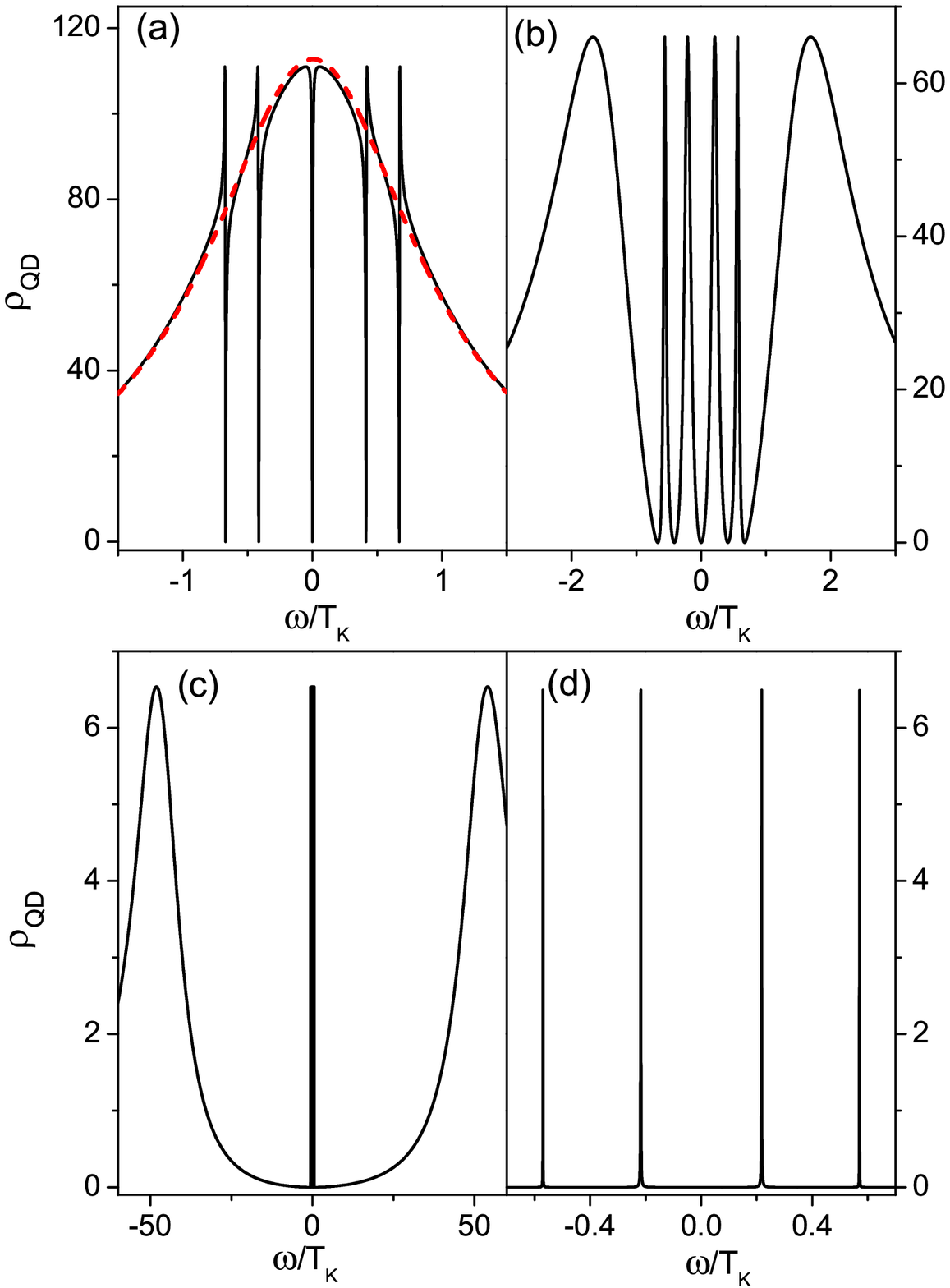}}}
\caption{(Color online) Density of states as a function of energy with magnetic flux $\Phi=0.25\Phi_0$. In the panel (a)   $t=0.1\, \xi$ (black solid  line) and  for disconnected QR, $t=0$ (read dashed line). In the panel (b)
$t=\xi$ and in the panel (c) $t=10\,\xi$. In the panel (d) it is shown in detail, the two central Loretzian peaks for  $t=10\,\xi$}
\label{figDOS2}
\end{figure}

\begin{figure}[tbp]
\centering
\rotatebox{0}{\scalebox{0.4}{\includegraphics{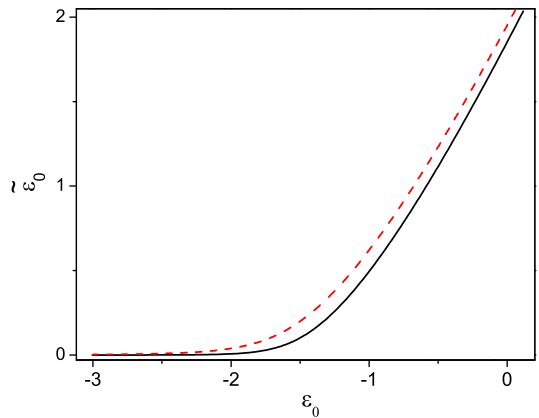}}}
\caption{(Color online) Renormalized energy vs gate voltage for weak coupling regime, $t=0.4\xi$ (black solid line) and strong coupling regime, $t=10\xi$ (red dashed line) for $\Phi=0$.}
\label{eo}
\end{figure}

In Fig.\ref{figDOS} (a), (b), (c) and (d)
we display the LDOS at the QD for weak, intermediate and strong
coupling regimes respectively for $\Phi=0$.
In the weak coupling regime the LDOS is slightly perturbed by the ring states appearing as Fano anti-resonances in an otherwise clear Kondo peak that results from the QD connected to its left and right leads.This behavior can be understood by inspection of equation \ref{G0ret}. The anti-resonances
appear at the ring state energies, which are the poles of $Q(\omega)$. When the interaction with the ring increases the original Kondo peak is now highly perturbed and the
DOS is now characterized by resonances that correspond to the poles of the Green's function and by two side bands with energies that get more distant from the
Fermi region as $t$ increases. These side bands correspond originally to the continuum Kondo peak that, splitted, is pushed away from the Fermi region. Well inside the strong regime in the Fermi region there are several discrete resonances of almost zero width while the continuum has been spread out from the Fermi region. This is clearly depicted in Fig.\ref{figDOS}(c) and (d). In this regime the impurity spin is completely screened by the spin of QR electrons near the Fermi level and there is no Kondo spin-spin correlation with the conducting electrons in the leads. It is very interesting to properly characterize these two regimes and the crossover between them. When the system is in the Kondo regime the renormalized local level of the QD, within the slave boson formalism, is fixed at the Fermi level independently of the gate potential applied to the QD, indicating the existence of a peak, the Abrikosov Shul resonance. This characteristic of the renormalized $\tilde{\varepsilon}_0$ is the finger print of the Kondo regime as far as the SBMFT formalism is concerned\cite{laercio}. In Fig. \ref{eo} we show the behavior of the  $\tilde{\varepsilon}_0$ as a function of $\varepsilon_0$. It is clear from the figure that independently of the system being in the weak or strong coupled regime the local level is renormalized to zero, when  $\varepsilon_0$ is below the Fermi energy. In the weak coupled regime, as it is clearly shown by the continuous resonance at the Fermi level of the DOS, the system is in the traditional bulk Kondo regime. In the strong coupling regime, the Kondo peak appears as a bunch of discrete levels very concentrated at the Fermi level  reflecting the fact that the system is in a Kondo like regime corresponding to an impurity connected to a quantum box, which has been named a Kondo box. As $\Phi=0$ does not give rise to semi-occupied state of the QR because no ring state coincides with the Fermi energy, the screening of the QD spin by the QR spins is possible through virtual excitations of one electron of the nearest to the Fermi energy double occupied ring state to the nearest non-occupied one \cite{Kbox}. This argument is not valid in presence of an external applied potential, if one of the level is tunned to be within the left and the right Fermi level in which case the state is semi-occupied and no virtual excitations are required to screened the QD spin.

We have a quantitative different situation  when $\Phi =0.25\Phi_0$ in which case the states in the ring are not degenerated and they are symmetrically distributes above and below the Fermi level with one of them, just at the Fermi level, occupied by only one electron as depicted in Fig. \ref{spectrum} and in the corresponding LDOS at the QD shown in Fig.\ref{figDOS2}. In this case the emergence of the Kondo box regime is easier, it is reached for smaller values of $t$ \cite{Kbox}. Within the scope of the SBMFT formalism this is the case because the renormalized energy $\tilde{\varepsilon}_0$ of the QD state interacts strongly with the QR state at the Fermi level, as both have the same energy, producing a bonding and anti-boding state as can be seen in  Fig.\ref{figDOS2}(d). The anti-bonding state will be double occupied by two opposite spin electrons. This is the mechanism that can explain the anti-ferromagnetic correlation between the QD spin and the spin of the electron populating the state at the Fermi level of the QR, which gives rise to the Kondo box regime.  However, as far as the essential physics we are analyzing is concerned, the behavior of the system in the strong coupling regime is only quantitatively weakly dependent on $\Phi$, as can be concluded by inspection of Fig.\ref{figDOS}, \ref{figDOS2}.

\begin{figure}[tbp]
\centering
\rotatebox{0}{\scalebox{0.35}{\includegraphics{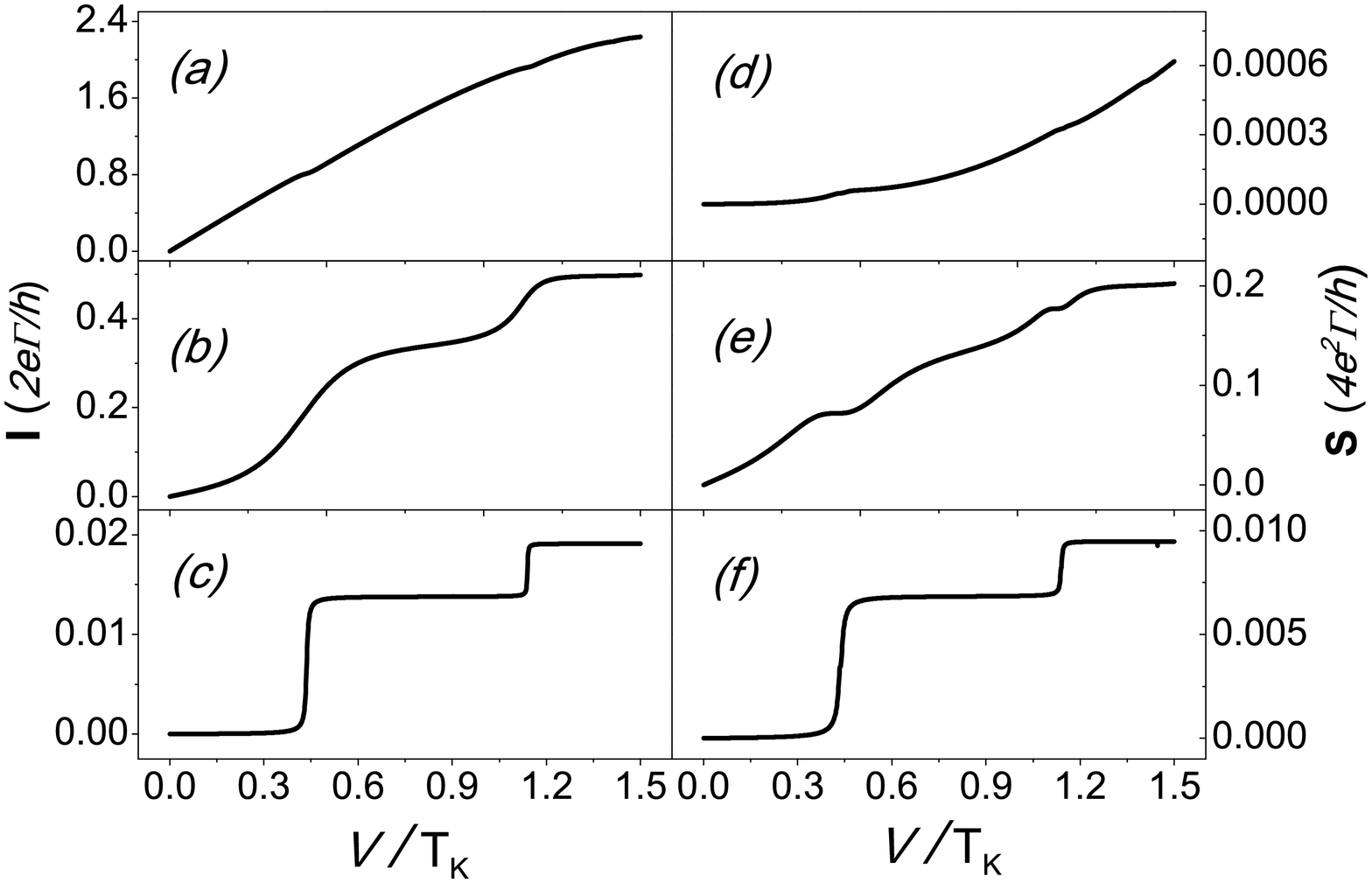}}}
\caption{Current (left panel) and Shot noise (right panel) versus bias voltage $V$, for magnetic flux $\Phi=0$, and
a) $t=0.1\,\xi$ , b) $t=\xi$
and c) $t=10\,\xi$.}
\label{JSvsV}
\end{figure}

\begin{figure}[tbp]
\centering
\rotatebox{0}{\scalebox{0.44}{\includegraphics{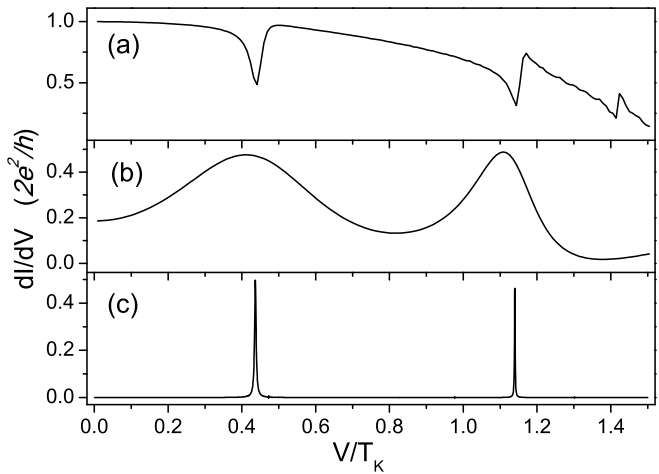}}}
\caption{Differential conductance as a function of applied voltage
for $\Phi=0$  and a) $t=0.1\,\xi$ , b) $t=\xi$
and c) $t=10\,\xi$.}
 \label{DiffGvsV}
\end{figure}

In order to adequately characterize the different regimes of the system, we calculate the current $I$ and the shot noise $S$ as a function of the potential bias $V$
for different coupling regimes. The Fig.\ref{JSvsV} displays $I$ ($S$) in left (right) panels restricted to the  $\Phi=0$ case. In the weak and
intermediate regime, the current and shot-noise increase smoothly
with the potential bias. However, in the strong coupling regime  $S$ and $I$ have steps for especial values of the external potential, which position depends upon $\phi$.  When the region between the left and right Fermi level includes a discrete local Kondo like peak, the current is able to flow increasing abruptly. Although this discrete states localized at the QD, as shown in  Fig.\ref{figDOS} and \ref{figDOS2}, are a resultant of the screening produced by the free spins in the QR, they provide a path for the electrons to go along, showing that these states have  a superposition with the leads. In the first plateau, the shot noise is $S = 2eI$ and in the last two $S= eI$ (Fig.\ref{JSvsV} (c) and (f)). The above results show that the measurement of the current and the noise permits to make a clear distinction and characterization of the regimes of the system. This is as well reflected in the differential conductance $dI/dV$, that is depicted in Fig.\ref{DiffGvsV} as a function of the bias.  In the weak coupling regime, we can see that the differential
conductance displays Fano anti-resonances any time that a level of the QR enter in the region between the left and the right Fermi energies. However in the strong coupling regime the differential conductance shows clear peaks corresponding to the steps in the current that were shown in  Fig.\ref{JSvsV}.

\begin{figure}[tbp]
\centering \rotatebox{0}{\scalebox{0.45}{\includegraphics{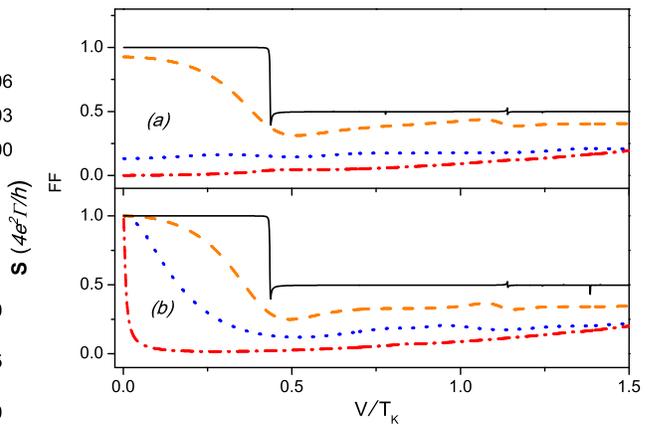}}}
\caption{(Color online). Fano factor versus bias, for two values of the magnetic flux, a) $\Phi=0$
and b) $\Phi=0.25\,\Phi_{0}$, for weak, intermediate and strong coupling regime, $t=0.2\,\xi$ (red dashed-dotted line), $t=0.6\,\xi$ (blue dotted line),
$t=\xi$ (orange dashed line) and (solid black line) $t=10\,\xi$}
\label{FFvsV}
\end{figure}

\begin{figure}[tbp]
\centering
\rotatebox{0}{\scalebox{0.4}{\includegraphics{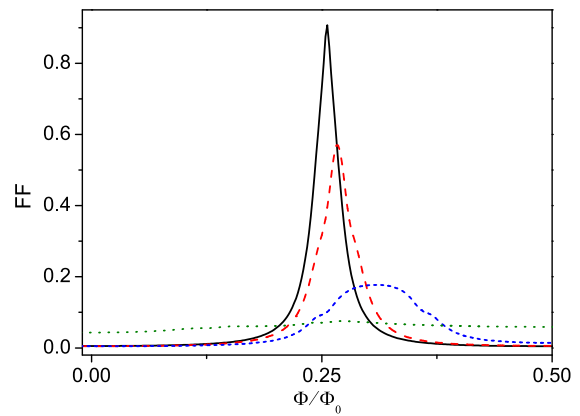}}}
\caption{(Color Online). Fano factor as function of the magnetic flux, for weak coupling regime ($t=0.4\, \xi)$ and different values of the bias. $V=0.02\, T_k$ (black solid line), $V=0.06\, T_k$ (read dashed line), $V=0.2\, T_k$ ( blue short dashed line), $V=T_k$ (greed dotted line) }\label{FF-FLUX}
\end{figure}

To complete the description of the transport properties, we study the Fano factor of the system.
Fig. \ref{FFvsV} displays the Fano factor as a function of
bias $V$, for two values of the magnetic flux, (a) $\Phi = 0$ and (b) $\Phi = 0.25$, for different QD-QR coupling in the range
$0.4\xi < t<10\xi$. In the strong
coupling regime the Fano factor have two plateaus and drops from $FF = 1$ to $FF = 1/2$ at $V \approx 0.4 T_{k}$. We
observe this behavior for various values of the magnetic flux, showing that the Fano factor is independent on the  flux in the Kondo box regime. On the other hand,  in the weak
coupling regime, $t=0.1\xi$, the Fano factor profile depends strongly on the
magnetic flux. As it is shown in Fig. \ref{FFvsV} (a) for $\Phi = 0$
and $V<<T_{k}$, the Fano-factor is $FF<<1$ and in Fig.
\ref{FFvsV} (b)  for $\Phi = 0.25 \, \Phi_0$, $FF\approx 1$.

In the strong coupling regime $S=2eI$ in the
first plateau, (see Fig.\ref{JSvsV}) and hence $FF=1$ and
in the last two plateaus $S=eI$ or equivalently $FF=1/2$.

This behavior is a consequence of the properties of the transmission probability $T(\omega,V)$ in the strong coupling regime. In the first
plateau $T(\omega,V)\approx 0$, then,
\begin{eqnarray}
[1-T(\omega,V)]\times T(\omega,V)\approx
T(\omega,V)
\label{approximation}
\end{eqnarray}
From Eqs.(\ref{current-shotnoise}) and (\ref{approximation}), it follows that in the first plateau
$S=2eI$.
In the Kondo box regime,  we can represent the transmission probability by a superposition of Briet-Wigner resonances.
\begin{eqnarray}
T(\omega,V)=\sum_{\alpha}\frac{\tilde{\eta}^{2}}{(\omega-\tilde{\varepsilon}_{\alpha})^{2}+\tilde{\eta}^{2}}
,\label{Briet-Wigner}
\end{eqnarray}
where $\tilde{\varepsilon}_{\alpha}$  and $\tilde{\eta}$ are the positions  and width of the Kondo box resonances, respectively. Then, from Eqs. \ref{current-shotnoise} and
\ref{Briet-Wigner} with $-V<\tilde{\varepsilon}_{\alpha}<V$, the
current $I$ and shot noise $S$ can be written as follows,
\begin{eqnarray}
I&=&\frac{4e}{h}M\tilde{\eta} \arctan\bigg(\frac{V}{\tilde{\eta}}\bigg)\nonumber\\
S&=&\frac{4e^{2}}{h}M\tilde{\eta}\frac{
\big[\big(\frac{V}{\tilde{\eta}}\big)^{2}+1\big]\arctan\big(\frac{V}{\tilde{\eta}}\big)-\frac{V}{\tilde{\eta}}}{\big(
\frac{V}{\tilde{\eta}}\big)^{2}+1}.
\end{eqnarray}
where $M$ is the number of resonances in the range of the applied bias. In the limit $V/\tilde{\eta}>>1$, $I\simeq 2\pi e\tilde{\eta}/h$ and $S\simeq 2
\pi e^{2}\tilde{\eta}/h$ or equivalently $S=eI$ and $FF=1/2$, in agreement with the
numerical results.

Finally, we calculate the Fano factor as a function of  the magnetic flux for different values of the bias, in the weak coupling regime ($t=0.4\,\xi$). As already discussed, the strong coupling regime is almost independent of $\Phi$. The results are depicted in Fig.\ref{FF-FLUX}. For small bias $V \ll T_{k}$, the Fano factor has a strong dependence on the magnetic flux, with a maximum around $\Phi =0.25\Phi_0$.  When the magnetic flux
is $\Phi\approx 0.25\Phi_{0}$ the side coupled QR, as it can be
observed in Fig.\ref{spectrum}, provides an state
of energy $\varepsilon_{n}(0.25\Phi_{0})=E_{F}=0$ that can be visited by an electron
which interferes destructively with the electron that goes directly through the QD. As a
consequence the current intensity decrease and the Fano factor reaches it's maximal allowed value,
$FF=1$.  Away of the maximum, all QR energy levels are far enough from the Fermi energy and the direct transport through QD predominates  without any destructive interference. Then the ratio
$S/2eI<<1$ and $FF\approx 0$.  By increasing the bias up $V=T_k$, the Fano factor becomes almost independent of the magnetic flux. As the Fano factor is almost independent of the magnetic flux in the strong coupling  Kondo box regime, as discussed above, the experimental measurement of this factor is another clear way of characterizing the system

We  demonstrate that the differential conductance, the shot noise and the Fano coefficients are all measurable properties that allow a very detailed characterization of the two Kondo regimes and the crossover region that the system goes through by manipulating its parameters.

\section{Summary}\label{Summary}

 In this work, we investigate the properties of a very interesting Kondo phenomena that results from the interplay between the traditional bulk Kondo effect and the so-called Kondo box regime. We  study the transport properties of a strongly correlated quantum dot attached to two leads and to a quantum ring (QR) pierced by a magnetic flux. In this system the QR  acts as a quantum box, which energies can be continuously modified by the application of the magnetic field. The crossover between these two regimes was studied by changing the connection of the QD to the leads and to the QR.

In the weak coupling regime, we have showed that the differential conductance develops a sequence of
Fano-Kondo antiresonances as a consequence of destructive interferences
between the $N$ discrete quantum ring levels with the conducting
Kondo channel provided by the leads. In the strong coupling regime the differential conductance has very sharp resonances when one of the Kondo discrete sub-level characterizing the Kondo box is tuned by the applied potential.
 We were able to demonstrate that the conductance, the current fluctuations and the Fano factor result to be the relevant physical magnitudes to be analyzed in the parameter space to reveal the physical properties of these two Kondo regimes and the crossover region between them.

The transport properties of this system are so extremely  dependent upon it parameters that it could have interesting potential applications as an active part of a nanoscopic circuit.

\section*{acknowledgments}
V.M.A and P.A.O acknowledges support from CONICYT PSD 65. M.P and P.A.O acknowledges support from FONDECYT program Grants
No. 1100560, and No. 1100672 and E.V. A.  acknowledges support from the brazilian financial agencies FAPERJ(CNS program) and CNPq (BP and CIAM).

\end{document}